\documentclass[12pt]{iopart}
\usepackage[latin9]{inputenc}
\usepackage{color}
\usepackage{amsbsy}
\usepackage{graphicx}
\usepackage{esint}
\PassOptionsToPackage{normalem}{ulem}
\usepackage{ulem}

\makeatletter
\usepackage{iopams}
\usepackage{iopams}

\usepackage{epsfig}\usepackage{epsf}\usepackage[babel=true,kerning=true]{microtype}\usepackage{soul}\usepackage{color}

\makeatother

\begin{document}

\title[Acoustic waves in a von-Kármán vortex street like
configuration of disclinations]{Propagation of acoustic waves in a von-Kármán vortex street like
configuration of disclinations}

\author{Sébastien Fumeron, Bertrand Berche}

\address{ Statistical Physics Group, IJL, UMR Université de Lorraine - CNRS
7198 BP 70239, 54506 Vandœuvre les Nancy, France}

\author{Fernando Moraes}

\address{Departamento de Física, CCEN, Universidade Federal da Paraíba,
Caixa Postal 5008, 58051-900, João Pessoa, PB, Brazil }

\author{Fernando Santos %
\thanks{On leave from: Departamento de Matemática Universidade Federal de
Pernambuco, 50670-901, Recife, PE, Brazil%
}}

\address{Wolfson Centre for Mathematical Biology, Mathematical Institute,
University of Oxford, OX1 3LB Oxford, U.K.}

\author{Erms Pereira}

\address{Escola Politécnica de Pernambuco, Universidade de
Pernambuco, Rua Benfíca, 455, Madalena, 50720-001 Recife, PE, Brazil }
\begin{abstract}
We show that  a simple model of 1D array of topological defects
in a crystalline environment is able to guide acoustic waves, depending
on the angle of the incident wave. We comment on a recently proposed geophysical mechanism explaining the mantle dynamics by the presence, in some crystalline materials there,  of wedge disclinations. 
\end{abstract}

\pacs{43.35.+d Ultrasonics, quantum acoustics, and physical effects of
sound - 91.60.Lj Acoustic properties - 91.30.Ab Theory and modeling,
computational seismology - 91.10.Kg Crystal movements and deformation}

\maketitle

\submitto{\JPCM}

\section{Introduction}

Wave propagation in elastic media is of prime interest in a great
number of problems in geophysics, ranging from seismology \cite{Sato98,Kaufman00}
to mineral physics \cite{Goebbels86,Turner94,Anugonda01}.
For example, the microstructural characterization of rocks can efficiently
be done from ultrasonic sounding: early works show that measurements
of wave speed or attenuation \cite{Mason47,Vary80}, as well as speckle
fields \cite{Goebbels80}, carry precious informations to retrieve
the inner structure of polycrystalline media where scattering effects
prevail. The Earth's crust consists in a set a rigid pieces, the plates,
which are moved by the deformations of an underlying layer, the mantle.
These plates rub against each other, causing disturbances (such as
earthquakes) that radiate seismic energy in the lithosphere. During
their propagation, high-frequency primary seismic waves are scattered
by randomly distributed heterogeneties, such as faults, irregular
topography of inner layers or space variations of rocks properties.
As a result of scattering events, the energy carried by ultrasonic waves
is known to obey the radiative transfer equation (RTE), that is a
Boltzmann equation that corresponds to a local balance on a radiometric
quantity, the specific intensity \cite{Margerin05,Weaver94}.

\begin{figure}[h!]
\centering \includegraphics[height=1.4in]{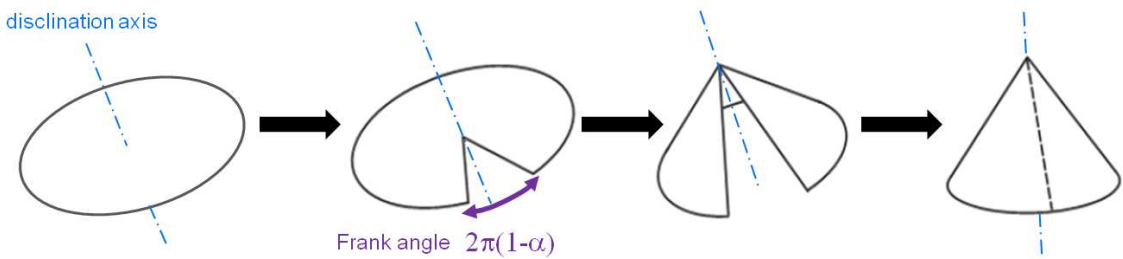} \caption{Volterra cut-and-glue process for a disclination.}
\label{discli} 
\end{figure}

In this geophysics context, the mantle dynamics and its ability to
deform was recently explained by the presence of wedge disclinations
inside olivine-rich rocks \cite{Cordier14,Hirth14}. This addresses
the challenging problem of ultrasound wave propagation and energy
concentration in the presence of topological defects, which is the
object of this article. The structure of a wedge disclination can
be easily understood from a Volterra "cut-and-glue" process: 1)
a cut is made in an elastic cylinder, 2) a wedge of material is removed
(positive Frank angle) or added (negative Frank angle) and 3) the
two edges of the cut are glued together. Fig.\ref{discli} shows the
Volterra process for the creation of a positive disclination in an
elastic disc. If the disc is allowed to buckle into the third dimension,
it becomes a cone which shares geometric properties with the disclinated
disc: they are both described by an otherwise flat metric with a curvature
singularity at the origin. Obviously, for real three-dimensional materials,
stress builds up since buckling is not allowed because it would have
to be into the fourth space dimension. Single disclinations are then energetically
unfavorable in 3D solid crystals because of their too high elastic
cost. 

However, that energy cost can be reduced in specific self-screening
configurations involving wedge disclinations dipoles: as considered
in \cite{Cordier14}, the self-screening reveals to be very efficient
for olivine presenting a periodic wedge disclination array of alternate
Frank angle signs.

In this work, we study the transport of scalar bulk waves in the presence
of an array of disclination dipoles. In the first part, we build a
geometric model for the distribution of disclinations that was considered
in \cite{Cordier14}, and the main properties of geodesic paths followed
by the waves are discussed. Then, we examine the foundations of the
radiative transfer equation for the specific intensity
from the standpoint of kinetic theory. Finally, the complete form
of the non-stationnary RTE in the presence of the defects is presented
and analyzed, allowing for a discussion of focusing/defocusing of
the energy in the system.

\section{The von Kármán vortex street geometry}

General relativity is the standard example where a physical phenomenon
(gravity) is represented by geometrical properties (curvature, torsion).
As testified by analogue gravity models \cite{Moraes00,Barcelo05},
the relevance of differential geometry does not restrict to gravitation,
but it can also be used for a wide range of classical and quantum
systems electromagnetic waves \cite{PBA02,metamats}, superfluid helium \cite{Volovik03},
2D electrons gas \cite{Vozmediano,Sinha}, liquid crystals \cite{Pereira13}...).
The pioneering works of Katanaev and Volovich \cite{Katanaev92} showed
the efficiency of this geometric approach to investigate the propagation
of elastic waves in the presence of topological defects such as screw
dislocations, edge dislocations or disclinations. Single disclinations
are line sources of curvature and they correspond to the generation
of a conical geometry, as shown in Fig.~\ref{discli}.

\begin{figure}[h!]
\centering 
\includegraphics[width=1\linewidth]{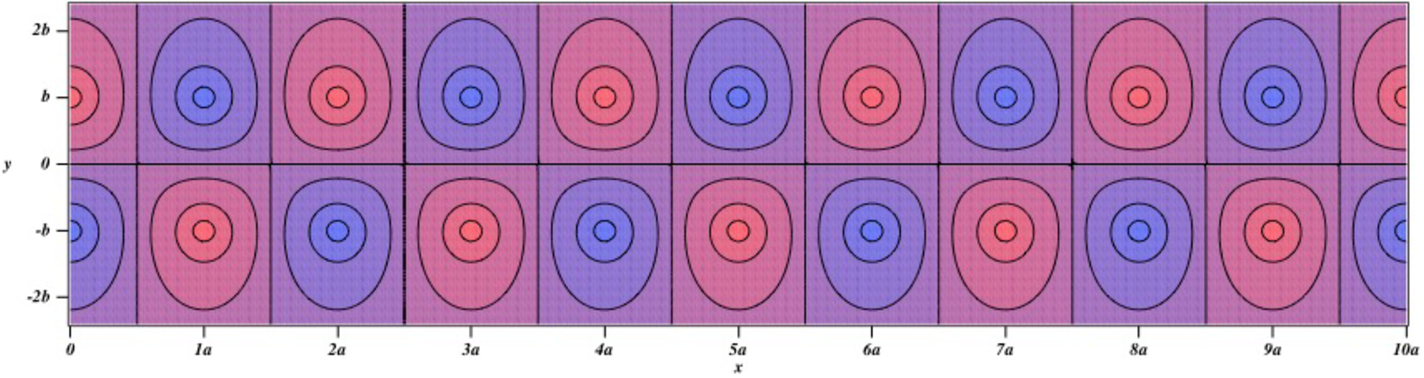} \vspace{-1cm}
\caption{A representation of the von Kármán vortex street distribution of alternate
disclinations given by the contour plot of the "potential" (\ref{fhat}).
\label{row}}
\end{figure}

In this section, we describe an array of disclination dipoles in terms
of differential geometry. Consider two rows made of an infinite number
of alternate disclinations separated by distance $2a$ and the rows
by $2b$. As illustrated in Fig.\ref{row}, the positive disclinations
(red contours) are at points $(na,(-1)^{n}b),n\in\mathbb{Z}$, while
negative disclinations (blue contours) have coordinates $(na,(-1)^{(n+1)}b),n\in\mathbb{Z}$.
\noindent 
Then, following a procedure given in \cite{Letelier01}, the corresponding
background geometry is generally described by the spacetime line element
\begin{equation}
ds^{2}=-c^{2}dt^{2}+e^{-4V(x,y)}\left(dx^{2}+dy^{2}\right)+dz^{2}=g_{\mu\nu}dx^{\mu}dx^{\nu}.
\end{equation}
$c$ would be the velocity of light in the general relativity
context, but here it stands for the local speed of wave packet, $g_{\mu\nu}$
is the metric tensor (Greek indices run from 0 (time index) to 3 and
the Einstein summation convention on repeated indices is used \cite{Schutz}).
In the function 
\begin{equation}
V(x,y)=(1-\alpha)f_{V}(x,y),
\end{equation}
$\alpha$ is related to the topological charge of a single disclination,
whereas $f_{V}$ describes the space distribution of the defects.
Such distribution of defects gives rise to a non-zero curvature tensor
component \cite{Letelier01}: 
\begin{equation}
R_{xyxy}=2(1-\alpha)e^{4(1-\alpha)f_{V}}(\partial_{xx}+\partial_{yy})f_{V}.\label{riem-tens}
\end{equation}

In the case of a single row of alternate defects at $y=b$ (positive
disclinations at $..-4a,-2a,0,2a,4a...$ and negative disclinations
at $..-3a,-a,a,3a...$), this function writes 
\begin{eqnarray}
\fl   f_{V}(x,y)  = \frac{1}{2}\sum_{p=-\infty}^{+\infty} \left\{ \ln\left[\left(x-2pa\right)^{2}+(y-b)^{2}\right]-\ln\left[\left(x-[2p-1]a\right)^{2}+(y-b)^{2}\right]\right\}\qquad \label{eqf} \\
  \fl \phantom{f_{V}(x,y) } = \frac{1}{2}\sum_{n=-\infty}^{+\infty} \left(-1\right)^{n}\ln\left[\left(x-na\right)^{2}+(y-b)^{2}\right] \label{sum-glob}
\end{eqnarray}
and the curvature becomes 
\begin{equation}
(\partial_{xx}+\partial_{yy})f_{V}\propto\sum_{p=-\infty}^{+\infty}\left[\delta(x-2pa)-\delta(x-(2p-1)a)\right]\delta(y-b).\label{riem-tens2}
\end{equation}
Introducing the complex variable $\xi=x+i(y-b)$, one obtains 
\begin{eqnarray}
2f_{V}(x,y) & = & \sum_{n=-\infty}^{+\infty}\left(-1\right)^{n}\ln\left[\left(\xi-na\right)\left(\bar{\xi}-na\right)\right]\label{fv1}\\
 & = & \ln\left(\xi\bar{\xi}\right)+\sum_{n=1}^{+\infty}\left(-1\right)^{n}\ln\left[\left(\xi^{2}-n^{2}a^{2}\right)\left(\bar{\xi}^{2}-n^{2}a^{2}\right)\right]\\
 & = & 2\hat{f}_{V}+4\sum_{n=1}^{+\infty}\ln(an),\label{fv1}
\end{eqnarray}
where $\bar\xi$ denotes the complex conjugate and the function $\hat{f}_{V}$ can be obtained by separating odd
and even contributions in (\ref{sum-glob}) 
\begin{eqnarray}
2\hat{f}_{V}(\xi) & = & \ln\left[\xi\prod_{m=1}^{+\infty}\left(1-\frac{\xi^{2}}{4m^{2}a^{2}}\right)\bar{\xi}\prod_{p=1}^{+\infty}\left(1-\frac{\bar{\xi}^{2}}{4p^{2}a^{2}}\right)\right]\nonumber \\
 &  & -\ln\left[\prod_{m=1}^{+\infty}\left(1-\frac{\xi^{2}}{(2m-1)^{2}a^{2}}\right)\prod_{p=1}^{+\infty}\left(1-\frac{\bar{\xi}^{2}}{(2p-1)^{2}a^{2}}\right)\right].\label{fv2}
\end{eqnarray}
From (\ref{riem-tens2}), it appears that the singular structure is
left invariant when adding a constant to $\hat{f}_{V}$. Therefore
(\ref{fv1}) and (\ref{fv2}) represent the same geometry up to a
scaling factor. By using the identities 
\begin{equation}
\sin(x)=x\prod_{m=1}^{+\infty}\left[1-\left(\frac{x}{m\pi}\right)^{2}\right],\;\;\;\cos(x)=\prod_{m=1}^{+\infty}\left[1-\left(\frac{2x}{(2m-1)\pi}\right)^{2}\right],
\end{equation}
it finally comes that 
\begin{eqnarray}
\hat{f}_{V}(x,y) & = & \ln\left|\sin\left(\frac{\pi}{2a}\left[x+i(y-b)\right]\right)\right|-\ln\left|\cos\left(\frac{\pi}{2a}\left[x+i(y-b)\right]\right)\right|\\
 & = & \frac{1}{2}\ln\left[\frac{\cosh^{2}\left(\frac{\pi}{2a}(y-b)\right)-\cos^{2}\left(\frac{\pi x}{2a}\right)}{\cosh^{2}\left(\frac{\pi}{2a}(y-b)\right)-\sin^{2}\left(\frac{\pi x}{2a}\right)}\right].
\end{eqnarray}
Therefore, when considering two lines of alternate disclinations as
in Fig. \ref{row}, the function $\hat{f}_{V}$ becomes: 
\begin{eqnarray}
\fl\hat{f}_{V}(x,y)=\frac{1}{2}\ln\left[\left(\frac{\cosh^{2}\left(\frac{\pi}{2a}(y-b)\right)-\cos^{2}\left(\frac{\pi x}{2a}\right)}{\cosh^{2}\left(\frac{\pi}{2a}(y-b)\right)-\sin^{2}\left(\frac{\pi x}{2a}\right)}\right)\left(\frac{\cosh^{2}\left(\frac{\pi}{2a}(y+b)\right)-\sin^{2}\left(\frac{\pi x}{2a}\right)}{\cosh^{2}\left(\frac{\pi}{2a}(y+b)\right)-\cos^{2}\left(\frac{\pi x}{2a}\right)}\right)\right].\ \label{fhat}
\end{eqnarray}
We can think of this function (in fact $V$(x,y)) as a sort of gravitational
"potential" acting on point masses that move in the presence of
the arrays of defects (see, for example, reference \cite{grats}).

\begin{figure}[h!]
\centering \includegraphics[height=9cm]{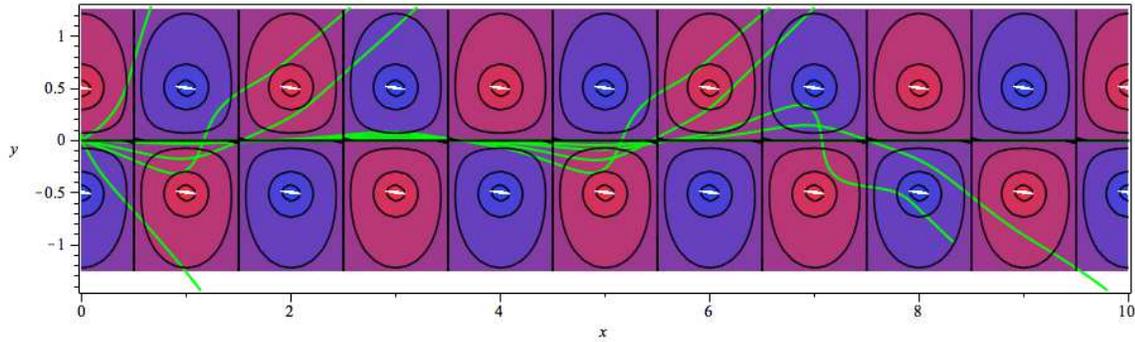} \vspace{-1cm}
 \caption{Different geodesics, shot from the origin, in the von Kármán vortex
street geometry. Depending on the shooting angle, the propagation
of the acoustic wave may be guided by the street of topological defects.}
\label{FigGeodesics1} 
\end{figure}

\begin{figure}[h!]
\centering \includegraphics[width=1\linewidth]{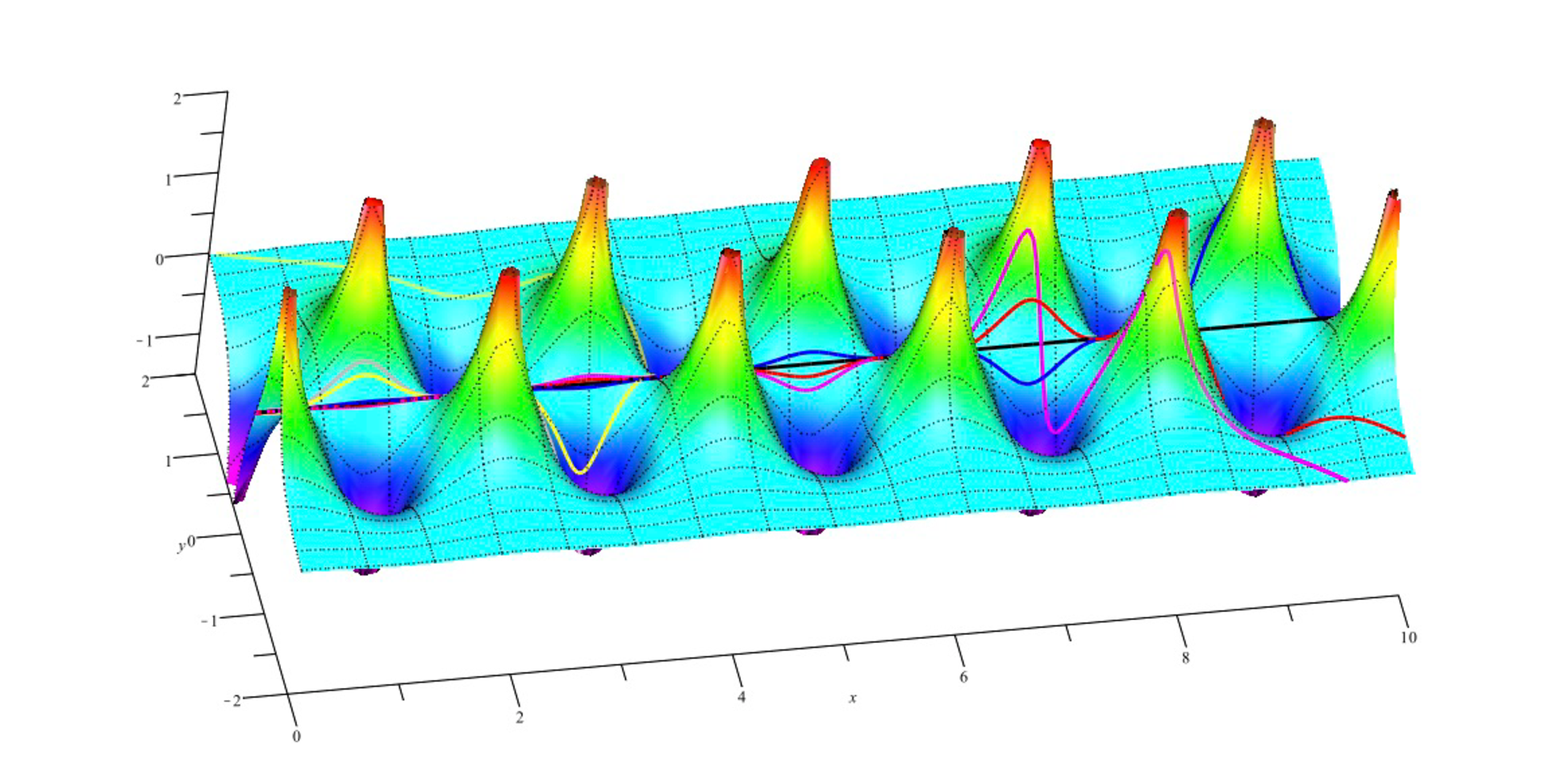} %\includegraphics[height=9cm]{teste.eps}
%\vspace{-1cm}
\caption{Geodesics in the von Kármán vortex street in the potential landscape.}
\label{Fig3dGeodesics} 
\end{figure}

As a consequence, the paths followed by acoustic waves are no longer
straight lines, but they are the geodesics of the von Kármán vortex
street geometry, that is trajectories of shortest
lengths in that geometry. They obey the so-called geodesic equations:
\begin{equation}
\frac{d^{2}x^{\mu}}{d\lambda^{2}}+\Gamma_{\rho\sigma}^{\mu}\frac{dx^{\rho}}{d\lambda}\frac{dx^{\sigma}}{d\lambda}=0,\label{geodesic}
\end{equation}
where $\lambda$ is an affine parameter along the path and $\Gamma_{\rho\sigma}^{\mu}$
are the Christoffel connections which can be expressed from the metric
tensor components as 
\begin{equation}
\Gamma_{\rho\sigma}^{\mu}=\frac{g^{\mu\nu}}{2}\left(\partial_{\rho}g_{\nu\sigma}+\partial_{\sigma}g_{\nu\rho}-\partial_{\nu}g_{\rho\sigma}\right)\label{christoLC}
\end{equation}
since there is no torsion. After some calculations, (\ref{geodesic})
reduce to \cite{Letelier01}: 
\begin{eqnarray}
 &  & \ddot{x}-2\left(\dot{x}^{2}-\dot{y}^{2}\right)\partial_{x}V-4\dot{x}\dot{y}\partial_{y}V=0,\\
 &  & \ddot{y}-2\left(\dot{x}^{2}-\dot{y}^{2}\right)\partial_{y}V-4\dot{x}\dot{y}\partial_{x}V=0.\label{geodesic2}
\end{eqnarray}
Although the geodesics' system of differential equations above are
analytical by nature, the inherent complexity of
$f_{V}(x,y)$ in Eq. (\ref{eqf}) leads us to cumbersome expressions
for the system in Eq. (\ref{geodesic2}). We circumvent such complexity
by using symbolic algebra, thus yielding a quite complex system of
differential equations for the geodesics, as illustrated in a flexible
Maple code, available from the authors upon request, for arbitrary
$\alpha$, $a$, and $b$. In order to get both quantitative and qualitative
informations about the geodesics in the von Kármán street, we systematically
solved the system of geodesic equations. We fixed $\alpha=1/2,a=1,b=0.5$
and solved numerically the resulting system of differential equations.
Without loss of generality, we chose geodesics starting at the origin,
$(x_{0}(0)=0,y_{0}(0)=0)$, with unitary initial ``speed'', $x'(0)=\cos(\theta_{0}),y'(0)=\sin(\theta_{0})$,
which defines the shooting angle. In Fig. \ref{FigGeodesics1} we
illustrate some geodesics in the von Kármán vortex street for a few
shooting angles. In Fig. \ref{Fig3dGeodesics} we depict some geodesics
in the potential landscape. 

\noindent 

The acoustic rays, as represented by the geodesics seen in Figs. \ref{FigGeodesics1}
and \ref{Fig3dGeodesics}, are very sensitive to the shooting angle.
Furthermore, they appear to be attracted by the positive defects while
repelled by the negative ones. This is quite natural,
since the former have the geometry of a cone (with less space than
the plane) and the latter of an anticone (with more space than the
plane). This makes the geodesic bend toward the positive defect and
away from the negative defect. In other words, the rays give a feeling on how acoustic propagation occurs in the medium with an array of defects. We can
infer that the defects may be used to design acoustic devices like
lenses or waveguides, which previously requires an accurate model to understand how energy is carried along these geodesics. This is the subject of
the following section.

\section{Radiative transfer equation and Clausius invariant}

Originally, radiative transfer consists in a phenomenological description
of the interactions of infrared electromagnetic waves with participating
matter. Based on the pioneering works by Khvolson \cite{Khvolson90},
Schuster \cite{Schuster05} and Schwarzschild \cite{Schwarzschild06},
this theory borrows concepts from radiometry (measurable quantity
such as the radiative flux), classical optics (Fermat principle to
get the optical paths) and quantum theory (Planck blackbody function
to account for the thermal emission by matter). In this section, we
introduce the essential concepts needed for the rest of the paper
and we remind the essentials of this phenomenological approach.

\begin{figure}[h!]
\centering \includegraphics[height=2in]{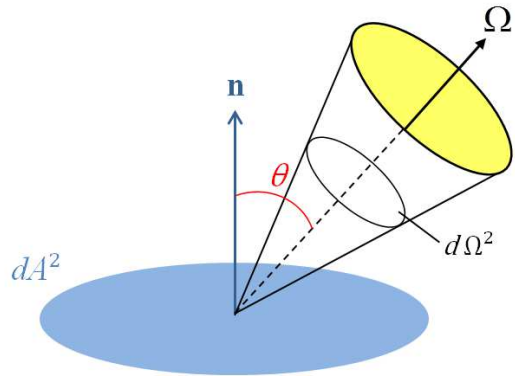} \caption{Definition of the specific intensity.}
\label{rad} 
\end{figure}

The central quantity is the specific intensity $I_{\nu}$ that corresponds
to the amount of energy crossing an element $dA^{2}$ of area during
time $dt$, in a frequency range between $\nu$ and $\nu+d\nu$, and
within an element $d\Omega^{2}$ of unit solid angle centered on direction
$\boldsymbol{\Omega}$ according to \cite{Modest03} (see Fig. \ref{rad}):
\begin{equation}
I_{\nu}=\frac{dE}{\cos\theta\: dt\: dA^{2}d\nu\: d\Omega^{2}}.\label{Iradio}
\end{equation}

The equation governing $I_{\nu}$ (known as the RTE) is derived from
a simple energy balance along any energy path: specific
intensity decays ought to absorption and scattering processes, and
it gets stronger because of scattering and inner source terms. In
its usual form, it is therefore given by \cite{Modest03,Chandra60}:
\begin{equation}
\frac{d}{ds}I_{\nu}\left(t,\mathbf{r},\mathbf{\Omega}\right)=-\left(\frac{1}{l}+\frac{1}{l_{a}}\right)I_{\nu}\left(t,\mathbf{r},\mathbf{\Omega}\right)+S\left(t,\mathbf{r},\mathbf{\Omega}\right),\label{RTE}
\end{equation}
where $d/ds$ is the total path length derivative, $l$ is the scattering
mean free path, $l_{a}$ is the absorption length, where both can
depend on the position. The source term writes 
\begin{equation}
S\left(t,\mathbf{r},\mathbf{\Omega}\right)=\frac{1}{l}\int_{4\pi}p\left(\mathbf{\Omega}',\mathbf{\Omega}\right)I_{\nu}\left(t,\mathbf{r},\mathbf{\Omega'}\right)d^{2}\Omega'+S^{0}\left(t,\mathbf{r},\mathbf{\Omega}\right).
\end{equation}
Here, $p\left(\mathbf{\Omega}',\mathbf{\Omega}\right)$ is the probability
that a beam propagating in direction $\mathbf{\Omega}$ is scattered
in direction $\mathbf{\Omega'}$ and $S^{0}$ is due to the radiative
energy emitted by sources embedded in the medium. This last term will
be neglected in the remainder of this work. This equation is relevant
for the present context, since it also applies for the acoustic waves
(compression, no shear) propagating in elastic media \cite{Weaver94,Margerin05}.

Despite its handiness, the phenomenological approach hides some intricate
points. Indeed, anytime a symmetry of physical properties of the participating
medium is lost, (\ref{RTE}) must be modified. As a matter of fact,
in nonstationnary and graded-index media, the adiabatic invariant
along a path is no longer the specific intensity but the so-called
Clausius invariant \cite{Kravtsov96}: 
\begin{equation}
C_{\nu}=\frac{I_{\nu}}{n(\bold{r})^{2}\nu^{3}},\label{clausius}
\end{equation}
where $n\left(\bold{r}\right)$ is the refractive index (the same
result was also found from other arguments by \cite{Marechal68,MTW73,Pomraning73,Fumeron08}).
It must be emphasized that (\ref{clausius}) also holds for radiative
transfer of acoustic waves, as an analog of Fermat principle can be
obtained in elastodynamics \cite{Horz95}. Hence, considering the
balance on Clausius invariant, the modified RTE now writes as \cite{Fumeron08}:
\begin{equation}
\frac{d}{ds}C_{\nu}=-\left(\frac{1}{l}+\frac{1}{l_{a}}\right)C_{\nu}+\frac{1}{l}\int_{4\pi}p\left(\mathbf{\Omega}',\mathbf{\Omega}\right)C_{\nu}d^{2}\Omega'.\label{MRTE}
\end{equation}
Integral solutions of (\ref{MRTE}) can be deduced formally by the
variation of parameters. One illustrates this point considering the
simple case of a steady-state regime 
\begin{eqnarray}
\frac{I_{\nu}}{n^{2}}\left(s,\mathbf{\Omega}\right) & = & \frac{I_{\nu}}{n^{2}}\left(s_{0},\mathbf{\Omega}_{0}\right)\exp\left(-\int_{s_{0}}^{s}\frac{ds'}{l(s')+l_{a}(s')}\right)\nonumber \\
 &  & +\int_{s_{0}}^{s}\tilde{S}\left(s',\mathbf{\Omega}\right)\exp\left(-\int_{s'}^{s}\frac{ds''}{l(s'')+l_{a}(s'')}\right)ds',\label{IMRTE}
\end{eqnarray}
with 
\begin{equation}
\tilde{S}\left(t,\mathbf{r},\mathbf{\Omega}\right)=\frac{1}{l}\int_{4\pi}p\left(\mathbf{\Omega}',\mathbf{\Omega}\right)\frac{I_{\nu}\left(t,\mathbf{r},\mathbf{\Omega'}\right)}{n(\bold{r})^{2}}d^{2}\Omega'.\label{source-ref}
\end{equation}
It is important to recall that in (\ref{MRTE})-(\ref{source-ref}),
the balance is performed along ray paths. Whether it is electromagnetic
or elastic beams \cite{Horz95}, the ray paths obey Fermat principle
which, as previously stated, requires to determine the refractive
index. In the next section, we introduce the distribution of disclination
dipoles from a geometric standpoint and extract from it the effective
refractive index experienced by the acoustic waves.

\section{Radiative transfer equation in the von Kármán vortex street geometry}

Based on the two preceding sections, we are now focusing
on the modified RTE in the presence of the array of disclinations
and determine some of its most remarkable properties. First, we briefly
present the general procedure to obtain the effective refractive index
from a diagonal metric. Consider a wave propagating along direction
$\boldsymbol{\Omega}=(\sin\hat{\theta}\cos\hat{\phi},\sin\hat{\theta}\sin\hat{\phi},\cos\hat{\theta})$
($\hat{\theta}$ and $\hat{\phi}$ denote the polar and azimuthal
angles) in a geometric background described by the metric tensor $g_{\mu\nu}$,
then the dispersion relation of a wave associated to 4-wavevector
$K^{\mu}=(\omega/c,k\sin\hat{\theta}\cos\hat{\phi},k\sin\hat{\theta}\sin\hat{\phi},k\cos\hat{\theta})$
writes as $K^{\mu}g_{\mu\nu}K^{\nu}=0$, which gives 
\begin{equation}
-\frac{\omega^{2}}{c^{2}}\left|g_{00}\right|+k^{2}\left(g_{11}\sin^{2}\hat{\theta}\:\cos^{2}\hat{\phi}+g_{22}\sin^{2}\hat{\theta}\:\sin^{2}\hat{\phi}+g_{33}\cos^{2}\hat{\theta}\right)=0,\label{a1}
\end{equation}
(all space components $g_{ii}$ are positive). On the other hand,
the dispersion relation of a wave in a dielectric medium at rest and
of refractive index $n$ is given by: 
\begin{equation}
-\frac{\omega^{2}}{c^{2}}n^{2}+k^{2}=0.\label{a2}
\end{equation}
Substituting $k^{2}$ from (\ref{a2}) into (\ref{a1}) defines an
effective refractive index 
\begin{equation}
n\left(\bold{r},\boldsymbol{\Omega}\right)=\sqrt{\frac{\left|g_{00}\right|}{g_{11}\sin^{2}\hat{\theta}\:\cos^{2}\hat{\phi}+g_{22}\sin^{2}\hat{\theta}\:\sin^{2}\hat{\phi}+g_{33}\cos^{2}\hat{\theta}}}.\label{a3}
\end{equation}
Generally speaking, this refractive index depends both on the position
in the medium through the radial vector $\bold{r}$ and the local
propagation direction $\boldsymbol{\Omega}$, but it does not exhibit
any dispersion.

\begin{figure}[h!]
\centering \includegraphics[width=1\linewidth]{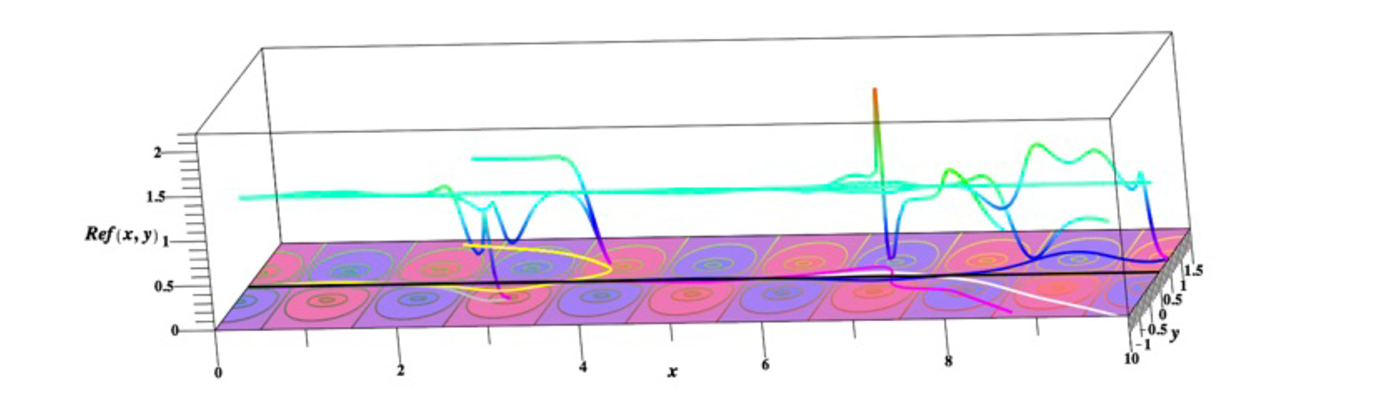} \vspace{-1cm}
 \caption{Refraction index projected on the von Kármán vortex street along some
representative geodesics. The refraction index is nearly constant
far from the defects and fluctuates around then. }
\label{FigGeodesics} 
\end{figure}

In the case of the von Kármán vortex street, the effective refractive
index writes as 
\begin{equation}
n(\bold{r},\hat{\theta})=\frac{1}{\sqrt{e^{-4V(x,y)}\sin^{2}\hat{\theta}+\cos^{2}\hat{\theta}}}.\label{effn}
\end{equation}
Notice that the refractive index depends both on the position $(x,y)$
and on the local propagation direction $\hat{\theta}$. In Fig. \ref{FigGeodesics}
it is presented a plot of $n(\bold{r},\hat{\theta})$ along selected
geodesics, showing strong fluctuations in the vicinity of the defects.

While the metric exhibits translational symmetry along the $z$ axis,
hence the coefficients $g_{\mu\nu}$ do not depend on $z$, this is
not true for the effective refractive index which looses this symmetry
through the dependence on the wavevector direction. From this perspective,
Eq.(\ref{effn}) might be misleading since only the angle $\hat{\theta}$
appears explicitly, but the presence of the azimuthal angle $\hat{\phi}$
is implicit in the $x,y$ dependence. From (\ref{clausius}), one
deduces the corresponding Clausius invariant 
\begin{equation}
C_{\nu}\left(t,\mathbf{r},\boldsymbol{\Omega}\right)=\frac{I_{\nu}\left(t,\mathbf{r},\boldsymbol{\Omega}\right)}{\nu^{3}}\left(e^{-4V(x,y)}\sin^{2}\hat{\theta}+\cos^{2}\hat{\theta}\right).
\end{equation}
Therefore, the modified RTE (MRTE) in the presence of the array of
defects is given by 
\begin{eqnarray}
\fl\frac{d}{ds}\left[\frac{I_{\nu}\left(t,\mathbf{r},\mathbf{\Omega}\right)}{\nu^{3}}\left(e^{-4V(x,y)}\sin^{2}\hat{\theta}+\cos^{2}\hat{\theta}\right)\right]=-\left(\frac{1}{l}+\frac{1}{l_{a}}\right)\frac{I_{\nu}\left(t,\mathbf{r},\mathbf{\Omega}\right)}{\nu^{3}}\left(e^{-4V(x,y)}\sin^{2}\hat{\theta}+\cos^{2}\hat{\theta}\right)\nonumber \\
+\frac{1}{l}\int_{4\pi}p\left(\mathbf{\Omega},\mathbf{\Omega}'\right)\frac{I_{\nu}\left(t,\mathbf{r},\mathbf{\hat{s}'}\right)}{\nu^{3}}\left(e^{-4V(x,y)}\sin^{2}\hat{\theta}+\cos^{2}\hat{\theta}\right)d^{2}\hat{s}'+S\left(t,\mathbf{r},\mathbf{\Omega}\right).\label{MRTE-a}
\end{eqnarray}

Some physical insights can be obtained by comparing Eq. (\ref{MRTE-a})
to the one without defects (\ref{RTE}). First, let us consider 
the simple steady-state case in a non-participating medium which writes
phenomenologically as 
\begin{eqnarray}
\frac{dI_{\nu}}{ds}=\left(2\frac{d}{ds}\ln\: n\right)I_{\nu}=-\frac{I_{\nu}}{l_{\alpha}}.\label{psBL}
\end{eqnarray}
Following \cite{Kravtsov96}, this defines a typical length which
measures the spatial inhomogeneities, $l_{a}$, given by 
\begin{eqnarray}
\frac{1}{l_{\alpha}\left(x,y,\hat{\theta}\right)} & = & \frac{2\sin\hat{\theta}}{\left(e^{-4V(x,y)}-1\right)\sin^{2}\hat{\theta}+1}\left[2\sin\hat{\theta}\: e^{-4V(x,y)}\left(\frac{\partial V}{\partial x}\frac{dx}{ds}+\frac{\partial V}{\partial y}\frac{dy}{ds}\right)\right.\nonumber \\
 &  & \left.\qquad+\cos\hat{\theta}\left(e^{-4V(x,y)}-1\right)\frac{d\hat{\theta}}{ds}\right].\label{inh}
\end{eqnarray}
The first derivatives $dx/ds$, $dy/ds$ and $d\hat{\theta}/ds$ are
obtained numerically from the geodesic equations (\ref{geodesic}).
There is no constraint that settles the signs of the different terms
involved in $l_{\alpha}$: hence, this parameter can be either of
positive or of negative sign, so that the array of defects either
damps or amplifies locally the specific intensity. That means that
the von Kármán vortex street does not only curve the geodesics, but
it also leads to a local geometric reduction/enhancement of energy
carried by the waves. Although (\ref{psBL}) is analog to a Beer-Lambert
law, it must be emphasized that there is no absorption process here.
The {\em inhomogeneity length} (\ref{inh}) rather describes the
local spatial enhancement (where it is negative) or local spatial
reduction (where it is positive) of the acoustic wave, like a focusing/defocusing
effect. 
Curvature doping of radiant intensity is well-known in radiative transfer
and was first discussed in \cite{PBA02}. This might be at the origin
of the deformation of the medium, where the energy transported by
the acoustic waves is locally amplified, allowing for a physical mechanism
for the mantle deformation due to the presence of topological defects,
as proposed in \cite{Cordier14}. 

\section{Concluding remarks}

To sum up, we proposed a simple toy-model for acoustic waves
propagation through a distribution of disclinations inside olivine-rich
rocks. In the presence of two rows of alternate wedge disclinations, two new phenomena arise. First, sound paths are curved by an effective background geometry, presenting alternate regions of positive and negative curvature originating from the alternate disclinations themselves. Second, as testified by the change in Clausius invariant, the energy carried by beams can be geometrically amplified/attenuated depending on the direction of the propagation. These two effects seem extremely sensitive to shooting conditions. 

Morevover, an asset of this formalism is that although polarization states of the waves and mode coupling
effects were not explicitly discussed, they can formally be incorporated
from the above results to obtain the generalization of the elastic
transfer equation. Investigating their impact on transfer will be the object of our further investigations.

\end{document}